\def\cm{{\rm\thinspace cm}}

\def\erg{{\rm\thinspace erg}}

\def\g{{\rm\thinspace g}}

\def\keV{{\rm\thinspace keV}}

\def\Msun{\hbox{$\rm\thinspace M_{\odot}$}}

\def\s{{\rm\thinspace s}}

\def\sr{{\rm\thinspace sr}}

\def\ergpcmsqpspsr{\hbox{$\erg\cm^{-2}\s^{-1}\sr^{-1}\,$}}
\def\ergpcmcups{\hbox{$\erg\cm^{-3}\s^{-1}\,$}}

\def\gpcm{\hbox{$\g\cm^{-3}\,$}}

\def\keVpcmsqpspsr{\hbox{$\keV\cm^{-2}\s^{-1}\sr^{-1}\,$}}

\def\psqcm{\hbox{$\cm^{-2}\,$}}

\documentstyle[psfig]{mn}
\begin{document}
\title{The mass density in black holes inferred from the X-ray background}
\author[]
{\parbox[]{6.in} {A.C.~Fabian and K. Iwasawa\\
\footnotesize
Institute of Astronomy, Madingley Road, Cambridge CB3 0HA \\}}

\maketitle
\begin{abstract}
The X-ray Background (XRB) probably originates from the integrated X-ray
emission of active galactic nuclei (AGN). Modelling of its flat spectrum
implies considerable absorption in most AGN. Compton down-scattering means
that sources in which the absorption is Compton thick are unlikely to be
major contributors to the background intensity so the observed spectral
intensity at about 30 keV is little affected by photoelectric absorption.
Assuming that the intrinsic photon index of AGN is 2, we then use the 30~keV
intensity of the XRB to infer the absorption-corrected energy density of the
background. Soltan's argument then enables us to convert this to a mean
local density in black holes, assuming an accretion efficiency of 0.1 and a
mean AGN redshift of 2. The result is within a factor of two of that
estimated by Haehnelt et al from the optically-determined black hole masses
of Magorrian et al. We conclude that there is no strong need for any
radiatively inefficient mode of accretion for building the masses of black
holes. Furthermore we show that the absorption model for the XRB implies
that about 85 per cent of accretion power in the Universe is absorbed. This
power probably emerges in the infrared bands where it can be several tens
per cent of the recently inferred backgrounds there. The total power emitted
by accretion is then about one fifth that of stars.
\end{abstract}


\section{Introduction}
Many current models for the X-ray Background (XRB) assume that it is due to
the summed emission from many Active Galactic Nuclei (AGN) with strong
intrinsic absorption (Setti \& Woltjer 1989; Madau et al 1994; Comastri et
al 1995; Celotti et al 1995). In this case most X-ray emission from AGN in the Universe must be
highly absorbed (Fabian et al 1998). Here we make a representative
correction of the spectrum of the XRB in order to deduce the current
radiation energy density of the XRB had absorption not been present. This is
converted to a bolometric radiation density using the X-ray to
bolometric ratio of a sample of unobscured AGN. Then from the simple
cosmology-free argument of Soltan (1982) we convert this energy density into
a mean mass density of black holes at the current epoch. Soltan (1982) took
the quasar counts in the optical B-band for his estimate so was only using
unobscured AGN. Our result exceeds his earlier estimate by a factor of about
7.5 and the upper limit of the more recent estimates of Chokshi \& Turner
(1992), which also use optical quasar counts, by a factor of 3. It agrees
with the very recent estimate of Salucci et al (1998), based on X-ray source
counts.

We show that the mean mass density of black holes is within a factor of two
of direct estimates based on the detection of black holes in nearby
galaxies. There is then no strong requirement for any significant mass build
up in black holes from some radiatively inefficient mode of accretion. Most
accretion power in the Universe is absorbed, and likely reradiated in the
infrared wavebands. This has implications for the IR background and source
counts.

\section{The mass density in Black Holes}

\begin{figure}
\centerline{\psfig{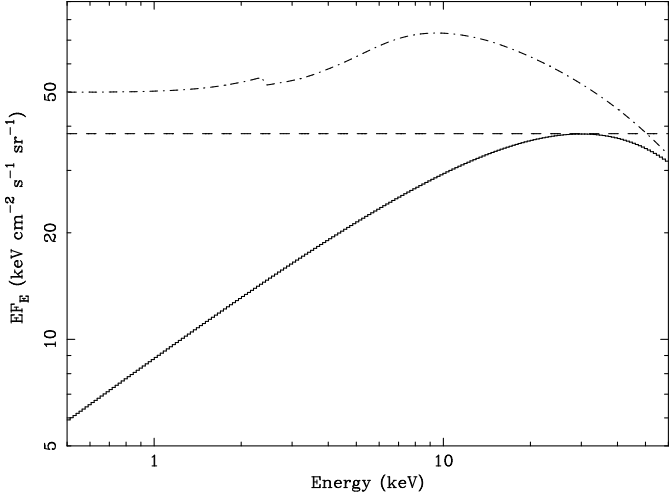}}
\caption{XRB spectrum (solid line) with the assumed unabsorbed spectrum of
photon index 2 (dotted line). A typical AGN spectrum with reflection, in
which the direct emission is a power law of photon index 2 with an
exponential cutoff of 300 keV is shown by the dot-dash line, matching around
the XRB peak. If unabsorbed quasars contribute 50 per cent of the XRB at 1
keV, then their contribution lies along the bottom of the figure.}
\end{figure}

We model the spectrum of the XRB as $$I_{\nu}=9 E^{-0.4}
\exp(-E/50\keV)\keVpcmsqpspsr \keV^{-1}.$$ This agrees with both the Marshall
et al (1980) result (HEAO A2 spectra taken in the 3--60~keV band) above
$\sim 7 \keV$ and that of Gendreau et al (1995) result (ASCA 0.5--7~keV
spectra) below that energy. The $E F_E$ spectrum then peaks at 30~keV with 
an intensity of $38.1 \keVpcmsqpspsr.$ 

The major assumption made in this paper is that the XRB spectrum above
30~keV is unaffected by absorption. This is because the likely maximum
photoelectric absorption for sources making a significant contribution to
the XRB intensity occurs at column densities of about $2\times
10^{24}\psqcm$, which has little effect above 30~keV in the rest frame, and
of course even less in redshifted sources. (We assume approximate Solar
metallicity for absorption purposes.) At higher column densities the Thomson
depth exceeds unity and Compton scattering significantly reduces the
transmitted power at all energies (see e.g. Madau et al 1994). Such
Compton-thick sources are therefore unlikely to play a major role in the
XRB, although they may be significant in number (perhaps one third of all
sources; Maiolino et al 1998).

We now assume that the intrinsic spectrum of the sources responsible for the
XRB has a photon index of 2 and matches the above spectrum at its $EF_E$
peak. A more complex spectrum including reflection can have a slightly
higher intensity (Fig.~1), particularly if redshifted. The caveats noted so
far mean that the true intrinsic spectrum is above the simple power-law
estimate, by some small factor.

The unabsorbed 0.66--3.33~keV (2--10~keV in the assumed mean restframe
redshift of 2, but with our adopted photon index this is independent of
redshift) intensity of the sources is $I_0=9.8\times
10^{-8}\ergpcmsqpspsr$. The unabsorbed radiation energy density is then
${\cal E_0}=4\pi I_0/c= 4.1\times 10^{-17}
\ergpcmcups.$  To convert this to an unabsorbed  bolometric radiation
density, we note that the mean ratio of the 2--10~keV to bolometric
luminosities for radio-quiet quasars in the compilation of spectral energy
distributions of Elvis et al (1994) is 3.3 per cent (we convert from the
tabulated monochromatic value of $L_{\rm X}$ to the 2--10~keV value by
multiplying by a factor of 1.6, using a photon index of 2; 6 out of every 7
objects have a luminosity ratio between 1 and 7.6 per cent). This is similar
to the ratio of 2 per cent found using Ferland's (1996) 'Table AGN' model,
which reproduces emission line ratios well. Denoting this fraction as
$(f/0.03)$, we find a total energy density of ${\cal E}'_0=4\pi I_0/fc=
1.2\times 10^{-15}(f/0.03)^{-1}\ergpcmcups.$ If we now assume, following
Soltan (1982), that this radiation has been produced by accretion at an
efficiency of 10 per cent (noting that (comoving) radiation energy density
decays as $1+z$ whereas mass density does not) we find a mass density
$$\rho=10(1+z){\cal E}'_0/c^2= 4.1\times 10^{-35}\gpcm.$$ This corresponds to
$6\times 10^{14}\Msun\ {\rm Gpc}^{-3}=6\times 10^{5}\Msun\ {\rm Mpc}^{-3}.$
A redshift of 2 is used here (see Miyaji et al 1998 for a discussion of
evolution models for X-ray observed AGN). 

This mass density is much greater than Soltan's original estimate and about
three times higher than that estimated by Chokshi \& Turner (1992) using
optical counts of unobscured quasars. It is about half that estimated by
Haehnelt et al (1998) from the product of the mean black hole mass to bulge
mass, estimated from spectroscopic data on the cores of nearby galaxies by
Magorrian et al (1998), and the mass density in galactic bulges (Fukugita et
al 1998). The Haehnelt et al estimate is likely to be fairly uncertain since
the method for black hole mass measurement of Magorrian et al may
overestimate the masses. Our result agrees with the very recent, and more
complicated, estimate of Salucci et al (1998) who use X-ray source counts
together with other factors. If a further 50 per cent of sources are
Compton-thick and, due to Compton down-scattering, contribute little to the
XRB spectral intensity, then our estimate rises proportionately.

An important conclusion of our result is that the absorption-corrected XRB
estimate and the more direct galaxy-core estimate of the local density of
massive black holes are within a factor of two. There is thus no strong need
to invoke very low efficiency accretion or any exceptional accretion flows.
What is required is that most accretion power is absorbed by surrounding
gas. This may require the geometries discussed by Fabian et al (1998). 

The fraction which is unabsorbed, in the hard X-ray spectrum, is about 6 per
cent of the total spectrum. Here we assume that all the optical and UV
emission are absorbed and that the infrared emission from AGN is already due
to reprocessing. The fraction is obtained by noting that only about one half
the 0.1--60~keV intrinsic X-ray spectrum is transmitted to make the XRB
spectrum (this is the ratio of the observed 0.1--60~keV intensity to that in
the inferred unabsorbed spectrum). The ratio of the 0.1--60~keV to 2--10~keV
fluxes for our assumed intrinsic spectrum is $r\sim4$, so if the 2--10~keV
flux is 3 per cent of the total flux of an AGN, the transmitted fraction of
the XRB, assuming the spectrum of Fig.~1, is $fr/2\sim6$ per cent.

The spectrum of the observed XRB does turn up below 1 keV in part due to
unabsorbed quasars and AGN, which from the integration of ROSAT source
counts (Hasinger et al 1993) contribute about half the total XRB intensity
at 1 keV. This corresponds to about 10 per cent of the assumed intrinsic
power-law spectrum so the total fraction of the accretion power emitted in
the Universe which escapes unabsorbed is therefore about 16 per cent. If we
further assume 50 per cent more AGN are Compton-thick with column densities
above $2\times 10^{24}\psqcm$ (cf. Maiolino et al 1998) then this final
number drops to about 12 per cent.

\section{The Infrared Background Light}

The absorbed intensity can now be used to estimate the IR background.
Assuming the absorber to be dusty gas, we have found that at least 85 per
cent of the emitted intrinsic flux for our adopted XRB spectrum is absorbed,
or an intensity $I_{\rm abs}=0.85 I_0/f\approx 3\times
10^{-6}\ergpcmsqpspsr.$ This is 3~nW~m$^{-2}$~sr$^{-1}$, for comparison with
the limits and detections derived of the infrared backgrounds derived and
reviewed by Hauser et al (1998) and Fixsen et al (1998). If much of the
absorbed flux emerges longward of 100$\mu$, which requires a significant
contribution occurring at redshifts much greater than 2, it will dominate
the FIR background and source counts (see also Almaini, Lawrence \& Boyle
1998). Wherever the reprocessed emission emerges, it corresponds to several
tens per cent of the likely background in the 10--100$\mu$m infrared bands
(Hauser et al 1998; Fixsen et al 1998) to most of the background longward of
$\sim200\mu$m, which is generally assumed to be dominated by the reprocessed
radiation from stars. This means that the integrated intrinsic radiation
from AGN is several tens per cent of that due to stars.

A check on this result is obtained by adapting an argument due to G.
Hasinger (priv. comm.). Magorrian et al (1998) find that the mass of a
central black hole $M_{\rm BH}$ is about $0.006$ times the mass of the host
bulge $M_{\rm b}$. If the present bulge mass is a fraction $a_{\rm b}$ of
the mass of the stars associated with that galaxy which have already burnt,
then the total radiation emitted over time is $0.0006 a^{-1}_{\rm b} M_{\rm
b} c^2$, where it is assumed that one tenth of a star undergoes nuclear
burning with an efficiency of 0.6 per cent. The total accretion energy
radiated from the black hole, at an efficiency of 0.1, is $0.1 M_{\rm BH}
c^2=0.0006 M_{\rm b} c^2$. If the history of star formation and accretion
are similar (see e.g. Boyle \& Terlevich 1998), then the contribution of AGN
to the background radiation density is $a_{\rm b}$ times that of stars. For
a Salpeter IMF extending to $0.4\Msun$, $a_{\rm b}\sim 0.2$. This fraction
can be increased if the radiative efficiency of the accretion is higher
(e.g. Kerr black holes) or the IMF is steeper than assumed, and decreased if
the IMF is flatter and a large non-bulge stellar component is included.

\section {Discussion and Conclusion}

We have estimated the local mean mass density in black holes from the
background light of AGN, correcting for the large absorption effects
necessary to make typical AGN explain the XRB spectrum. Our result of
$6-9\times 10^{5}\Msun\ {\rm Mpc}^{-3}$ (the higher value includes a
correction for Compton-thick sources) is within a factor of 2 to 1.5 of that
estimated from direct optical studies of the cores of nearby galaxies. The
range of uncertainty on our mass density estimate, based only on variations
in the adopted spectral energy distributions for AGN, is also about a further
factor of two.

The key assumptions are that a) the background above 30~keV is unaffected by
absorption and b) the typical spectrum of high redshift, and also of
absorbed AGN, are similar to that of the unobscured AGN used by Elvis et al
(1994) in their compilation of spectral energy distributions. The first
assumption is robust to reasonable changes in the relevant energy (e.g.
going to 40 or 50~keV in Fig. 1 would have little effect on the final
result). There is insufficient information to assess the second assumption,
but we do note that it is plausible that the intrinsic X-ray emission region
is unaffected by the absorption which is usually assumed to occur at much
greater radii from the central black hole. Finally, the total X-ray
intensity is reduced by about 30 per cent if the photon index drops to 1.5,
but the contribution of reflection in all cases increases our estimate (by
about 10 per cent for a photon index of 2).

We are concerned that the simple use of the Elvis et al (1994) spectral
energy distribution for AGN leads to the UV emission being counted twice.
This is because the infrared emission may be dominated by absorbed UV (and
soft X-ray) emission (assuming that we view the UV source direct, but the
absorbing matter which reradiates UV as infrared covers other lines of
sight). If we omit the infrared emission from that distribution the
bolometric luminosities drop by about one third. This increases $f\sim 0.05$
and decreases our mass density estimate by one third. It also means that the
total absorbed fraction of the power drops to about 80 per cent.

In conclusion, the spectrum of the XRB provides a robust estimate of the
total radiation emitted by AGN in the Universe. The agreement of the local
black hole density resulting from the conversion of the implied energy
density to mass with direct estimates from other wavebands indicates that
there is no obvious need for some radiatively inefficient form of accretion.
Most of the accretion power in the Universe has been absorbed and presumably
reradiated in the infrared.

\section{Acknowledgements}

We thank Omar Almaini, Guenther Hasinger and Andy Lawrence for discussions
and Elihu Boldt for comments. ACF thanks the Royal Society for support.

\end{document}